\documentclass[twocolumn,showpacs,preprintnumbers,amsmath,amssymb]{revtex4}
\usepackage{graphicx}
\usepackage{dcolumn}
\usepackage{bm}
\usepackage{epsfig}

\begin{document}

\title{Theory of microphase separation of
homopolymer--oligomer mixtures }

\author{Alexander Olemskoi}
\email{olemskoi@ssu.sumy.ua}
\author{Ivan Krakovsky}
\email{ivank@kmf.troja.mff.cuni.cz}
\author{Alexey Savelyev}
\email{alexsav@kmf.troja.mff.cuni.cz}
\affiliation{Department of Physical Electronics,
Sumy State University, Rimskii-Korsakov St. 2, 40007 Sumy, Ukraine\\
Department of Macromolecular Physics, Charles University\\
V Hole\v{s}ovi\v{c}k\'{a}ch 2, 180~00 Prague 8, Czech Republic}

\date{ \today }

\begin{abstract}

Microphase separated structure consisting of the periodic
alternation of the layers of stretched homopolymer chains
surrounded by perpendicularly oriented oligomeric (surfactant)
tails is studied for the systems with strong ionic) as well as
weak (hydrogen) interaction. Our approach is based on the fact
that the structure period is determined by alternating
associations between the head groups of the oligomer molecules and
interacting groups of the homopolymer chains. Distribution of
oligomers along the homopolymer chains is described by the
effective equation of motion with the segment number playing the
role of time. As a result, experimentally observed temperature
dependence of the structure period, as well as the dependence of
the point of order--disorder transition are determined as
functions of the oligomeric fraction.

\end{abstract}

\maketitle

\section{Introduction}

Various mesomorphic structures can be prepared  using  the strong
(ionic)  or weak (hydrogen) bonding, respectively, between
homopolymers and head--functionalized oligomers or surfactants
\cite{1}. The interaction between the head group of the oligomeric
molecule and suitable groups on the homopolymer chain, on the one
hand, and unfavorable polar--nonpolar interactions between the
non--polar tail of the oligomer and the rest of the system, on the
other one, can cause the microphase separation. The latter effect
results in a periodic alternation of the layers of stretched
homopolymer chains surrounded by perpendicularly oriented
oligomeric tails as it is shown in Figure 1.
\begin{figure}[htb]
\epsfig{file=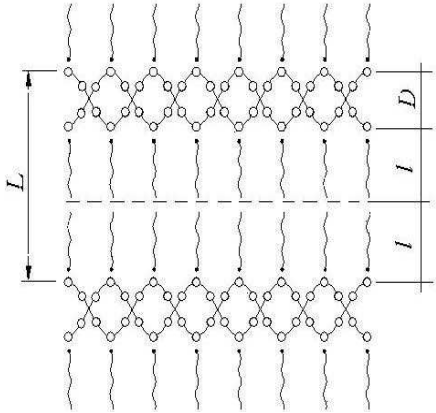,height=60mm,width=59mm,bbllx=0mm,bblly=0mm,bburx=129mm,bbury=111mm}
\caption{Schematic of the polymer--surfactant model for the
oligomeric fraction $x=1/3$.}
\end{figure}
An example of the ionically bonded system is represented by the
homopolymer-oligomer mixture of atactic poly(4-vinyl pyridine)
(P4VP) and dodecyl benzene sulfonic acid (DBSA) where due to a
very strong interaction the microphase separation is realized over
the whole temperature region \cite{2}. The peculiarity of the
systems of this kind is an increasing long space period $L$ as a
function of the oligomeric fraction $x$ defined as the number of
the oligomeric (DBSA) molecules per one pyridine ring (see Figure
2).
\begin{figure}[htb]
\epsfig{file=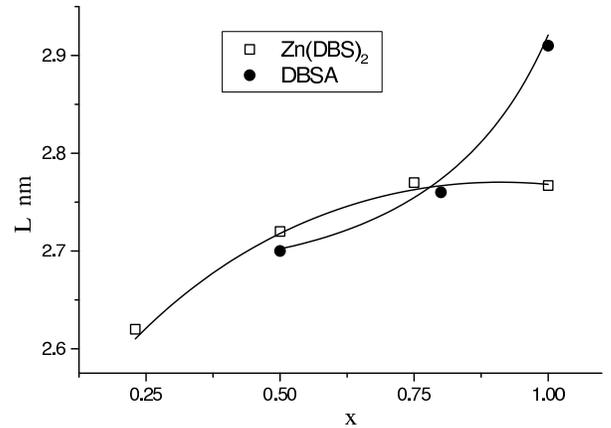,height=80mm,width=59mm,bbllx=23mm,bblly=24mm,bburx=190mm,bbury=253mm,angle=-90}
\caption{Long space period in strongly bonded systems as a
function of  the oligomeric fraction $x$. The solid lines
represent the dependencies obtained by fitting according to
Eq.(\ref{26}).  Experimental data obtained for  P4VP-(DBSA)$_x$
($\bullet$) and P4VP-(Zn(DBS)$_2$)$_x$ ($\square$) at room
temperature are taken from the Ref. \protect\cite{1}}
\end{figure}
More complicated behavior is inherent in the hydrogen bonded
systems where the weak interaction causes an order--disorder
transition to homogeneous high--temperature state \cite{1,3}. An
example of the system of this kind represents the mixture
P4VP--(PDP)$_x$  of the same homopolymer P4VP with the surfactant
3-pentadecyl phenol (PDP). Here, contrary to the ionically bonded
systems, the long space period decreases with increasing $x$ (see
Figure 3).
\begin{figure}
\epsfig{file=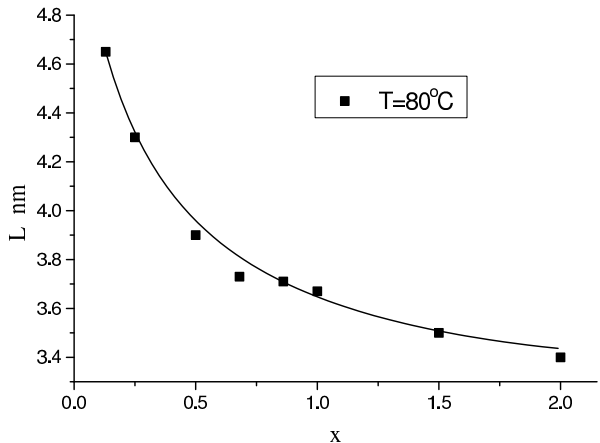,height=80mm,width=59mm,bbllx=23mm,bblly=24mm,bburx=190mm,bbury=253mm,angle=-90}
\caption{Long space period in the weakly bonded system as a
function of  the oligomeric fraction $x$. The solid line
represents the dependence obtained by fitting according to
Eq.(\ref{21}).  Experimental data obtained for  P4VP-(PDP)$_x$
($\blacksquare$) at temperature $T=80^o {\rm C}$ are taken from
the Ref. \protect\cite{3}}
\end{figure}
As it is shown in Figure 2, an intermediate behavior
(non--monotonous $x$--dependence of the long space period) was
found  in  the system P4VP-(Zn(DBS)$_2$)$_x$ where the ionic
interaction between  zinc dodecyl benzene sulfonate Zn(DBS)$_2$
and P4VP is somewhat weaker than in the P4VP-(DBSA)$_x$
system\cite{4}.

Principally important for our considerations is the decreasing
character of the temperature dependence of the long period found
experimentally for all the systems mentioned above \cite{1} ---
\cite{4}. However, such a character of the dependence appears in
the hydrogen bonded systems only within a finite temperature
interval limited by the temperatures of crystallization (or glass
transition $T_g$) from below and of the order--disorder transition
$T_c$ from above (see Figure 4).
\begin{figure}[htb]
\epsfig{file=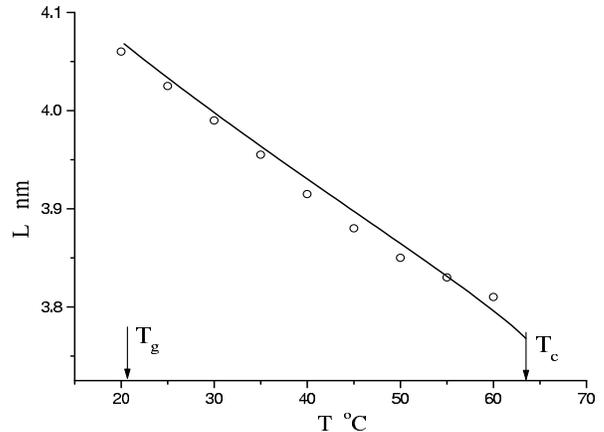,height=80mm,width=59mm,bbllx=16mm,bblly=14mm,bburx=107mm,bbury=129mm,angle=-90}
\caption{Temperature dependence of the long space period in the
weakly bonded system. The solid line represents the dependence
obtained by fitting according to Eq.(\ref{21}).  Experimental data
obtained for  P4VP-(PDP) at $x=0.85$ ($\circ$) are taken from the
Ref. \protect\cite{3}}
\end{figure}
As it is shown in Figure 5,  an increase of the oligomeric
fraction leads to a non--monotonous dependence  of the temperature
$T_c$ with a maximum occuring near  $x=0.85$.
\begin{figure}[htb]
\epsfig{file=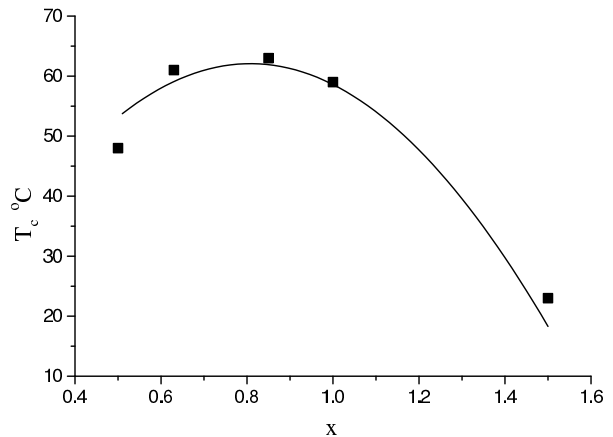,height=80mm,width=59mm,bbllx=23mm,bblly=24mm,bburx=193mm,bbury=256mm,angle=-90}
\caption{Order-disorder transition temperature, $T_c$,  for the
weakly bonded system as a function of the oligomer fraction $x$.
The solid line represents the dependence obtained by fitting
according to Eq.(\ref{16}). Experimental data obtained for
P4VP-(PDP)$_x$ ($\blacksquare$)  are taken from the Ref.
\protect\cite{3}}
\end{figure}
The  aim of this paper  is to explain peculiarities of the
microphased separated  homopolymer--oligomer mixtures  with both,
strong and weak bonding, within a framework of the unified scheme
of stochastic systems \cite{5}.

Our consideration is based on the obvious equality for the long
period $L=2 l+D$ where $l$ is the oligomeric chain length and $D$
is the thickness of the homopolymer layer being fixed due to
sharing of the surfactant molecules by homopolymer chains (see
Figure 1). Physically, this share is reduced to the inverse
magnitude $2\pi/\omega$ of the circular frequency of the
alternation of the oligomer heads along the homopolymer chain.
Then, the long period  can be  expressed by the following equation
\cite{1}
\begin{equation}
L=2l+D_0\omega^{-1},\quad
D_0\equiv2\pi\left(\chi^{1/6}n_0^{-1/3}\right)b\sim b \label{1}
\end{equation}
where $\chi\sim 10/N$ is the Flory interaction parameter ($N\sim
10^4$ is the degree of polymerization of the homopolymer chains),
$n_0\sim 10$ is the number of segments in the oligomer chain, $b$
is the segment length. The  frequency $\omega$ will be found from
the effective equation of motion which is stochastic in nature and
where the segment number $n$ ($n\le N$) plays the role of imagined
time \cite{6}.

\section{Basic equations}

The problem under consideration is addressed by the definition of
effective law of motion $c(n)$ that determines the sequence of the
alternation of oligomers along the homopolymer chain by means of
the occupation number,  $c(n)$, being $c(n)=1$ if the oligomer is
attached to the segment $n$, and $c(n)=0$ otherwise. In the limit
$N\to\infty$, the argument $n$ may be considered as a continuous
one, and the behavior of the system is governed by the action and
the dissipative functional
\begin{equation}
S=T\int\limits_0^N L\big(c(n),\dot c(n)\big){\rm d}n,\quad
R=\frac{\Theta}{2}\int\limits_0^N\big(\dot c(n)\big)^2{\rm d}n
\label{3}
\end{equation}
with dimensionless Lagrangian
\begin{equation}
L=K-\Pi;\qquad K\equiv\frac{m}{2}\big(\dot c(n)\big)^2,\quad
\Pi\equiv\frac{\tau}{2}\big(c(n)\big)^2. \label{4}
\end{equation}
Here dot denotes the derivative with respect to the segment number
$n$, $T$  is temperature in energy units, $\Theta$, $m$ and $\tau$
are inverse kinetic coefficient, inhomogeneity and interaction
parameters, respectively. The key point in our consideration
consists in that  the effective mass $m$  in Eq.(\ref{4})  is a
fluctuating parameter with the mean value $\bar m$ and the
variance $\overline{(m-\bar {m})^2}\equiv\sigma^2$ (bar denotes
the average, as usually). Then, after the averaging of the
exponent $\exp(-S/T)$ over Gaussian distribution of the bare mass
$m$, we obtain renormalized inhomogeneity energy
\begin{equation}
K=\frac{1}{2}\left({\bar m}-{\tilde\Delta}\right) \big(\dot
c(n)\big)^2,\ \ {\tilde\Delta}\equiv\frac{\sigma^2}{2}
\int\limits_0^N\big(\dot c(n')\big)^2{\rm d}n'. \label{7}
\end{equation}
As a result, taking into account the stochastic source $\zeta$,
relevant Euler equation arrives at non--linear Langevin equation
\begin{equation}
{\tilde m}\ddot c+n_c\dot c+\tau c = \zeta \label{8}
\end{equation}
where the effective mass ${\tilde m}={\tilde m}\{c(n)\}$,
characteristic number of correlating segments $n_c$ and
$\delta$--correlated noise $\zeta(n)$, respectively,  are
introduced in accordance with definitions:
\begin{eqnarray}
&{\tilde{m}}\equiv{\bar m}-{\tilde\Delta},\quad
n_c\equiv{\Theta\over T},&
\label{9}\\
&\langle\zeta(n)\rangle=0,\quad
\left\langle\zeta(n)\zeta(n')\right\rangle=\delta(n-n').&
\label{10}
\end{eqnarray}

To linearize the equation of motion (\ref{8}) within the
self--consistent approach, it is necessary to replace the
fluctuational term $\tilde\Delta$ of the renormalized mass in Eq.
(\ref{7}) by the averaged expression
\begin{equation}
\Delta=\sigma^2 \int S(\nu)\nu^2\frac{{\rm d}\nu}{2\pi} \label{11}
\end{equation}
where $S(\nu)\equiv\left\langle|c(\nu)|^2\right\rangle$  is the
structure factor in  the frequency representation (angle brackets
denote average over noise $\zeta$). Then, the Green function
$G\equiv \left\langle\delta c/\delta\zeta\right\rangle$ and the
structure factor $S=|G|^2$ take the forms:
\begin{equation}
G=\left[\tau(\nu)-{\rm i}n_c\nu\right]^{-1}, \quad
S=\left[\tau^2(\nu)+n_c^{2}\nu^2\right]^{-1} \label{12}
\end{equation}
where renormalized interaction parameter is introduced by
{\bf equalities}
\begin{eqnarray}
\tau(\nu)\equiv\tau-m_{ef}\nu^2,\quad m_{ef}\equiv{\bar m}-\Delta.
\label{13}
\end{eqnarray}

\section{Determination of the period of microphase structure}

To obtain physically observable values, at first one has to
determine the effective mass $m_{ef}$ given by Eqs.(\ref{13}),
(\ref{11}). On the basis of the theory of residues, using  the
structure factor,  Eq.(\ref{12}) arrives at the expression for
renormalization mass parameter
\begin{equation}
\Delta=\frac{\sigma^2}{n_c m_{ef}}. \label{14}
\end{equation}
Inserting here Eqs.(\ref{9}), (\ref{13}), we obtain the effective
inhomogeneity parameter
\begin{equation}
m_{ef}=\mu(T){\bar m},\quad
\mu\equiv\frac{1}{2}\left(1+\sqrt{1-\frac{T}{T_c}}\right)
\label{15}
\end{equation}
where the temperature domain is bounded above by the
characteristic temperature
\begin{equation}
T_c\equiv\left(\frac{\bar m}{2\sigma}\right)^2\Theta. \label{16}
\end{equation}
According to Eq.(\ref{15}) the effective mass decreases
monotonously with increasing temperature from the bare magnitude
${\bar m}$ at $T=0$ to ${\bar m}/2$ at $T=T_c$.

The divergency condition of the Green function (\ref{12}) arrives
at the proper frequency of the oligomer alternation
\begin{equation}
\nu_0=-{\rm i}\varpi\pm\omega \label{17}
\end{equation}
with imaginary and real parts
\begin{equation}
\omega^2\equiv\omega_0^2-\varpi^2,\quad
\omega_0^2\equiv\frac{\tau}{m_{ef}},\quad
\varpi\equiv\frac{n_c}{2m_{ef}}. \label{18}
\end{equation}
Insertion of Eqs.(\ref{9}), (\ref{15}) gives the temperature
dependence of the proper frequency
\begin{equation}
\omega^2=\frac{{\bar\omega}_0^2}{\mu^2}\left(\mu-
\frac{T_0^2}{T^2}\right) \label{19}
\end{equation}
where the bare frequency ${\bar\omega}_0$ and another
characteristic temperature $T_0$ are introduced as follows:
\begin{equation}
{\bar\omega}_0^2\equiv\frac{\tau}{\bar m},\quad
T_0\equiv\frac{\Theta}{2\sqrt{{\bar m}\tau}}. \label{20}
\end{equation}
As a result, combination of Eqs.(\ref{1}), (\ref{19}) leads to
the final result
\begin{equation}
L=2l+\frac{\mu(T)}{\sqrt{\mu(T)-T_0^2/T^2}}L_0 \label{21}
\end{equation}
where the characteristic length, $L_0$, is defined as
\begin{equation}
L_0\equiv\frac{D_0}{{\bar\omega}_0}\sim \sqrt{\frac{\bar
m}{\tau}}~b. \label{22}
\end{equation}

\section{Discussion}

The behavior of the system is determined by the relation
\begin{equation}
\kappa=\frac{T_c}{T_0}\equiv\frac{\sqrt{{\bar m}\tau}}{2}
\left(\frac{\bar m}{\sigma}\right)^2\geq 1 \label{23}
\end{equation}
where the minimal magnitude $\kappa=1$ fixes the choice of the
theory parameters according to the condition: $\sigma\leq
2^{-1/2}{\bar m}^{5/4}\tau^{1/4}$. It would seem  from
Eq.(\ref{23}) that the decrease of the temperaure $T_{c}$ with
weakening interaction when passing from the ionically bonded
system (such as P4VP-(DBSA)$_x$) to the hydrogen bonded one (e.g.,
P4VP--(PDP)$_x$) is caused only by the growth of the fluctuation
parameter $\sigma$ with respect to the  mean magnitude of the
inhomogeneity parameter $\bar m$. It appeares, however, that the
main reason for such behavior is a decrease of the
mean--geometrical magnitude $\sqrt{{\bar m}\tau}$ of the principle
coefficients in the generic Lagrangian (\ref{4}).

To clarify this problem, assume the three--parametric
$x$--dependencies for the  above parameters
\begin{equation}
\bar{m}=m_0+Ax(x_m - x),\quad \tau=\tau_0+Bx(x_\tau - x)
\label{25}
\end{equation}
with positive constants $m_0$, $\tau_0$, $A$, $B$, $x_m$,
$x_\tau$. In limiting case,  $\kappa\gg 1$ relevant to the
ionically bonded systems, when $m_{ef}\simeq{\bar m}$,
$\mu(T)\simeq 1$, the temperature $T_c$ is so large that the
temperature dependence (\ref{21}) takes the form
\begin{equation}
L=2l+\frac{L_0}{\sqrt{1-T_0^2/T^2}},\quad L_0\sim\sqrt{\frac{\bar
m}{\tau}}~b. \label{26}
\end{equation}
Then, fitting of the experimental data shown  in Figure 2
according to Eq.(\ref{26}) arrives at the following results. The
system P4VP-(DBSA)$_x$: $m_0=49$, $A=21$, $x_m=0.5$; $\tau_0=1$,
$B=1.1$, $x_{\tau}=0.5$; $b=1$ nm; $l=10$ nm. The system
P4VP-(Zn(DBS)$_2$)$_x$: $m_0=18$, $A=55$, $x_m=1.8$; $\tau_0=1$,
$B=0.1$, $x_{\tau}=2$; $b=1$ nm; $l=10$ nm. Respectively, for
$x=1$ Eq.(\ref{23}) gives values $\kappa=10^3,\ 10^2$ at ratios
$\sigma/{\bar m}\approx 0.045,\ 0.203$.

A much more complicated situation occurs  in hydrogen bonded
system P4VP--(PDP)$_x$, where the reduction of the parameter
(\ref{23}) leads to a more narrow temperature domain $T_0\div
T_c$. As a result, the temperature dependence of the period $L(T)$
obtains a more complicated form (\ref{21}) which keeps the
decreasing character as shown in Figures 3 and 4. The dependence
$L(x)$ at a fixed temperature may be described by the estimation
\begin{equation}
L\approx 2l+\sqrt{\bar m/\tau}~b \label{27}
\end{equation}
following from Eq.(\ref{21}) and conditions $\mu\approx 1$,
$T_0\ll T$. Then, the experimental data shown  in Figure 3 give
the following constrains: $\sqrt{m_0/\tau_0}=33.5$,
$B/\tau_0=1.6$, $x_{\tau}=3.1$. Respectively, the comparison of
the experimental data shown in Figure 4 with the fitting results
using Eqs.(\ref{15}), (\ref{16}), (\ref{20}) and (\ref{21}) yields
$\Theta/\sqrt{m_0\tau_0}=1517$, $l=8.7$ nm. Finally, Figure 5 and
Eq.(\ref{16}) give  $A/m_0=0.16$, $x_m=1.6$, $\Theta
m_0^2/4\sigma^2=276$.

As a result, taking $m_0=1$ at $x=1$ the magnitudes $A=0.16$,
$\Theta=5.5$, $\sigma=0.07$ are obtained and they provide the
value of $\tau_0$  smaller than  $10^{-3}$ which arrive at the
parameter $\kappa\approx 6.7$ and ratio $\sigma/{\bar m}=0.069$.
It is worth to mention that the value of
 the parameter $\tau_0< 10^{-3}$ is extremally small that
guarantees the validity of the Flory--Huggins approach.

To conclude, the model desribed above can explain successfully all
peculiarities obtained experimentally for various
homopolymer--oligomer mixtures with interaction of different
strength.

{\bf{Acknowledgement}} In this work, financial support by the
Grant Agency of the Czech Republic (grant GA\v{C}R 203/02/0653) is
gratefully acknowledged.

\end{document}